\documentclass[twocolumn,showpacs,preprintnumbers,amsmath,amssymb]{revtex4-1}

\usepackage{graphicx}
\usepackage{dcolumn}
\usepackage{bm}

\begin{document}


\title{Self-assembly of Brownian motor by reduction of its effective temperature.}
\author{Alexander Feigel}
\email{sasha@soreq.gov.il}
\author{Asaf Rozen}
\affiliation{%
Soreq NRC,\\
Yavne 81800, Israel}


\date{\today}

\begin{abstract}
Emergence, optimization and stability of a motor-like motion in a fluctuating environment are analyzed. The emergence of motion is shown to be a general phenomenon. A motor converges to the state with the minimum of effective temperature and with the corresponding minimum in the rate of conformation changes similarly as some stochastic processes converge to the states with minimum diffusion activity. This mechanism is important to bacterial foraging (chemotaxis). This work, therefore, raises an analogy between chemotaxis and the emergence of living-like systems. The implications include the deviation of stable natural or artificial machines from the minimum entropy production principle, with a novel self-assembly mechanism for the emergence of the first molecular motors and for mass fabrication of the future nanodevices.
\end{abstract}

\maketitle

\affiliation{%
Soreq NRC,\\
Yavne 81800, Israel}


\date{\today}

\maketitle

Analysis of microscopic ratchets is the focus of modern statistical physics. Energy and conformation fluctuations drive small systems away from macroscopic physical laws. The ratchet-like systems might rectify these fluctuations to generate mechanical propulsion or complex robotic functions\cite{Smoluchowski1912, Feynman2011, Hanggi2009}. This type of microscopic machines amuse us with behavior that resembles living systems, such as steady state operation when out of thermal equilibrium and with a seeming reduction of entropy.

A macroscopic ratchet includes a gear with asymmetric teeth and a pawl (see Fig. 1(a)). Asymmetry of the teeth and of the pawl only permits omni-directional rotation of the gear.  Motion of the teeth switches the pawl between “open” and “closed” states. The pawl in its open state allows a single tooth of the gear to pass in the allowed direction, then immediately switches to the closed state that prevents backward motion. A ratchet is a common part of mechanical jacks and precise clocks because its backward motion is forbidden.

A microscopic ratchet in a fluctuating environment might operate as a Brownian motor that converts thermal fluctuations into directed motion\cite{Smoluchowski1912, Feynman2011}. The implementation of this motor used to be a puzzle because a microscopic ratchet, contrary to a macroscopic one, might rotate in both directions. Fluctuations might keep the pawl in its open state and at the same time drive the gear in any direction.
Bidirectional motion of a ratchet at a micro scale is intrinsic because omnidirectional rotation in a uniform temperature environment resulting from thermal fluctuations might reduce entropy and, therefore, breaks the second law of thermodynamics.

A series of works showed the essential conditions for a realistic Brownian motor to operate\cite{Feynman2011, magnasco1993forced, Astumian1997, Ajdari1992, Hanggi2009, Reimann2002}. The generation of mechanical work when out of thermal fluctuations requires spatial asymmetry with a temperature gradient, a time-dependent modulation or another form of energy flux. This mechanism is considered a leading physical model for molecular motors\cite{Bustamante2001, Julicher1997, Mogilner1998, Astumian2010, Peskin1993} and inspired various accomplishments using the transport of particles in nanopores, in cold atoms lattices, in microfluidic circuits, and in superconducting and quantum nanodevices (for review see \cite{Hanggi2009} and corresponding references).

The question of stability and the closely related problem of the emergence of the Brownian motor, to the best of our knowledge, was not addressed yet. The stability of the Brownian motor is important to take analogy with microscopic ratchets, molecular motors and living system beyond the mere operation principle toward emergence and optimization. For the sake of the stability problem, one should consider fluctuations in basic properties of the ratchet, such as asymmetry, rather only in the position of its mechanical parts. For instance, what are the required conditions to ensure development of asymmetry with the corresponding motor function? Continual machinelike behavior is a common property of organisms. This question, therefore, addresses the fundamental problems of emergence and stability of living systems.

Here we show a universal mechanism for self-assembly of a Brownian motor. This mechanism is based on the reduction of the motor's effective temperature because of the emergence of spatial asymmetry and associated motion. The steady state of this self-assembly differs from the normally accepted minima whether by free energy\cite{Whitesides2002, Grzybowski2009} or entropy production\cite{Prigogine1967}. The proposed self-assembly permits the emergence and optimization of the motor function, rather than its mere emergence as a side effect of another optimization process. Self-assembly of a Brownian motor is an analogy of emergence of the first biological machines during pre-Darwinian chemical evolution\cite{Calvin1969}, a method for mass production of artificial nanomachines\cite{Whitesides2002} and a framework for analysis of fundamental questions about the stability of physical systems when out of thermal equilibrium.

The Smoluchowski-Feynman\cite{Smoluchowski1912,Feynman2011} ratchet is, probably, the most famous implementation of the Brownian motor (see Fig. 1(a)). The ratchet consists of two symmetric and nonsymmetric connected paddle wheels, kept at various temperatures $T_1$ and $T_2$. These temperature gradients and asymmetry cause the ratchet to move continually until the temperatures equilibrate. No continual motion occurs if all parts are symmetric, though temperatures still equilibrate through the ratchet's fluctuations. Temperature difference might be replaced by interaction with granular gas\cite{Eshuis2010}, cold atoms\cite{Renzoni2005} or even by interaction with living species\cite{DiLeonardo2010}.

Triangulita is an elegant implementation of one-dimensional Brownian ratchet\cite{VandenBroeck2004} (see Fig. 1(b)). This motor includes only two symmetric (rectangle) and asymmetric (triangle) heads. Triangulita has a linear average velocity if its head as a triangle breaks the symmetry along the axis of propagation and if there is a temperature difference between the heads of the ratchet. The various temperatures of triangle and rectangle parts correspond to the interaction with ideal gases. The temperatures and densities of the gases are $(T_1,\rho_1)$ and $(T_2,\rho_2)$ correspondingly. This model allows analytical treatment at the limit $m/M<<1$, where $m$ is the mass of the gas particles and $M$ is the mass of Triangulita.
\begin{figure}
    \begin{tabular}{c c}
      \multicolumn{1}{l}{{\bf\sf A}} & \multicolumn{1}{l}{{\bf\sf B}} \\
      \resizebox{0.25\textwidth}{!}{\includegraphics{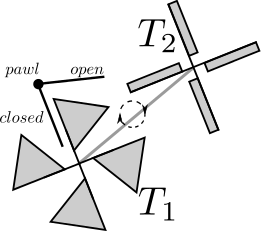}} &
      \resizebox{0.25\textwidth}{!}{\includegraphics{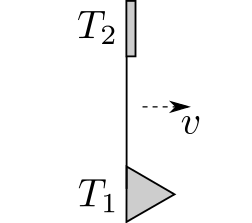}} \\
    \end{tabular}
    \caption{({\bf A}) Smoluchowski-Feynman Brownian motor. The motor consists of two heads connected by a common axis. The heads are asymmetric ratchet and symmetric paddle wheel. The motor rotate its heads are immersed in gases at different temperatures $T_1$ and $T_2$ correspondingly. Rotation continues until the temperatures equilibrate. This type of motor was the subject of several paradoxes and puzzles during the previous century. The ratchet is a physical model for artificial nanodevices and molecular motors such as kinesin. ({\bf B}) Triangulita Brownian motor. Similar to the Smoluchowski-Feynman motor, connected asymmetric triangle and symmetric rectangle propulse linearly if they are coupled to the various temperatures. Triangulita enables analytic analysis of the major characteristics of Brownian motors.}
    \label{fig1}
\end{figure}

To study self-assembly of Triangulita, we make a single modification to the model: the triangle part can revolve around the joint point (see Fig. 2). This modification facilitates the transition from the symmetric state with no motion to a position with maximum asymmetry and maximum propulsion velocity. The mechanism of rotation is chosen to be coupled to the effective temperature of the motor $T_{eff}$ instead of one of the reservoirs. This assumption is true if rotation mechanism is hidden inside the structure of Triangulita.

Triangulita with rotating triangle part might converge to an asymmetric state with maximum velocity by mutations that depend on effective temperature of the motor. Let us assume that the rotation of the triangle is slow compared with its linear motion and that triangle's orientation $\alpha$ changes by discrete steps $\Delta\alpha$ in clockwise or counterclockwise directions (see Fig. 2). Step in either direction is a transition through effective potential barrier $\Delta E$. In this case, the transition rate is\cite{Allen2006}:
\begin{eqnarray}
  \label{eq:rate1}
 R\propto \exp(-\Delta E/(k_bT_{eff}(\alpha)), 
\end{eqnarray}
where $k_b$ is Bolzman constant. Effective temperature $T_{eff}$ is:
\begin{eqnarray}
  \label{eq:teff3}
  T_{eff}(\alpha)\propto <v^2>-<v>^2,  
\end{eqnarray}
where $v$ is the ratchet's velocity. Effective temperature $T_{eff}$ depends on triangle's orientation $\alpha$ because Triangulita's velocity $v$ is a function of $\alpha$. Dynamics of Triangulita in $\alpha$ space then described by continuity equation: 
\begin{eqnarray}
  \label{eq:cont1}
  \frac{\partial P(\alpha)}{\partial t} = \frac{\partial}{\partial \alpha} J_{\alpha}(\alpha),
\end{eqnarray}
where $J_{\alpha}(\alpha)$ is the effective flux between the states with several orientations of the triangle. The steady distribution $P_{st}(\alpha)$ corresponds to zero flux $J_{\alpha}(\alpha)=0$. The flux depends on the rate (\ref{eq:rate1}) and, therefore, is a function of effective temperature.
\begin{figure}
\resizebox{0.5\textwidth}{!}{\includegraphics{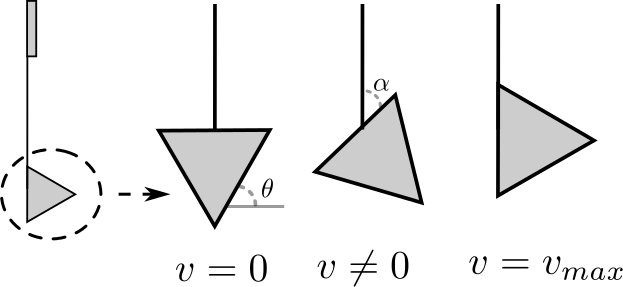}}
\caption{Self-assembly of Triangulita Brownian motor. Motion emerges with transition of triangle part from symmetric ($\alpha=\pi/2$ or $\alpha=\pi/6$) to asymmetric position ($\alpha=0$ or $\alpha=\pi/3$). Motion depends on the geometric form factors $<\sin^n(\theta)>$, where $\theta$ is an angle between tangential and a horizontal lines. In this work, this transition occurs because of reduction in effective temperature of the motor. Effective temperature affects the mutation rate of triangle's orientation. The motor converges to an asymmetric state with minimum effective temperature and maximum propulsion because the corresponding diffusion process leads the system toward the state with minimum effective diffusion coefficient. Chemotaxis of bacteria, for instance, occurs because of similar phenomena.}
\label{fig2}
\end{figure}

Some mutation processes converge to the states with minimum mutation activity\cite{schnitzer1993theory,lanccon2001drift}. For instance, in the case of constant step $\Delta\alpha$ with varying transition rate (\ref{eq:rate1}), the flux in the space of $\alpha$ is:
\begin{eqnarray}
  \label{eq:jtheta1}
  J_{\alpha}=\frac{\partial D(\alpha) P(\alpha)}{\partial \alpha},
\end{eqnarray}
where $P(\alpha)$ is a probability of the rotation state $\alpha$. The corresponding diffusion coefficient is\cite{schnitzer1993theory}:
\begin{eqnarray}
  \label{eq:DTbm}
  D(T_{eff})\propto \exp{-\frac{\Delta E}{k_b T_{eff}}}.
\end{eqnarray}
This flux results in a steady probability $P_{st}(\alpha)$:
\begin{eqnarray}
  \label{eq:rho1}
  P_{st}(\alpha)\propto \frac{1}{D(T_{eff}(\alpha))}.
\end{eqnarray}
The ratio of the probabilities to be in two states with different effective temperatures $T_{eff1}$ and $T_{eff2}$ is: 
\begin{eqnarray}
  \label{eq:rhorat1}
  \frac{P_{st}(T_{eff1})}{P_{st}(T_{eff2})}\propto\exp\left (\Delta E\left ( \frac{1}{T_{eff1}}-\frac{1}{T_{eff2}}\right )\right ).
\end{eqnarray}
The system, therefore, tends to concentrate at the state with the minimum effective temperature. The degree of concentration depends on the difference in the temperatures and value of the barrier $\Delta E$.

To derive effective temperature of Triangulita as a function of its orientation $\alpha$ let us summarize the required analytical results\cite{Meurs2004}. Triangulita Brownian motor is described by Boltzmann-Master equation:
\begin{eqnarray}
  \label{eq:bme1}
  \frac{\partial P(v,t)}{\partial t}=\int dv'\left [ W(v|v')P(v',t)-W(v'|v)P(v,t)\right ],
\end{eqnarray}
where $P(v,t)$ is probability density of velocity $v$ and $W(v|v')$ is transition rate from $v'$ to $v$. The motor changes its velocity because of interaction with the gas particles. In each thermal bath $i$, the interaction of gas with the heads of the motor is a function of the gas temperature $T_i$ and density $\rho_i$ with the size and shape of the motor's head. The latter is described by perimeter $S$ with the following form factors: 
\begin{eqnarray}
  \label{eq:sinn}
  <sin^n(\theta)>_i=\int_0^{2\pi}d\theta F_i(\theta)sin^n(\theta),
\end{eqnarray}
where $\theta$ is a tangent of the surface measured counterclockwise from the axis of motion and $F(\theta)d\theta$ is the fraction of the surface between $\theta$ and $\theta+d\theta$ (see Fig. 2). The lowest order velocity of Triangulita is:
\begin{eqnarray}
  \label{eq:vtriang}
&&<v>=\\\nonumber
&&=\sqrt{\frac{m}{M}}\sqrt{\frac{\pi k_bT_{eff}^{scale}}{8M}}\frac{\sum\rho_iS_i\left ( \frac{T_i}{T_{eff}^{scale}}-1\right )<sin^3(\theta)>}{\sum\rho_iS_i\sqrt{\frac{T_i}{T_{eff}^{scale}}}<sin^2(\theta)>}+\\\nonumber
&&+O\left (\left(\frac{m}{M} \right )^{\frac{3}{2}} \sqrt{\frac{1}{M}}\right ),
\end{eqnarray}
where scaling temperature is:
\begin{eqnarray}
  \label{eq:Teffsc1}
  T_{eff}^{scale}=\frac{\sum\gamma_iT_i}{\sum\gamma_i},
\end{eqnarray}
with friction coefficients:
\begin{eqnarray}
  \label{eq:gammai1}
  \gamma_i = 4S_i\rho_i\left (\frac{k_b T_im}{2\pi} \right )^{\frac{1}{2}}<sin^2(\theta)>_i.
\end{eqnarray}
Finally, the motion of Triangulita is described by the moments of its velocity $<v^n>$ expended as a series of small parameter $\epsilon=\sqrt{m/M}$.

Contribution of triangle part to the effective temperature (\ref{eq:teff3}) of Triangulita follows from eqs. (39) and (B2) of the work\cite{Meurs2004}:
\begin{widetext}
\begin{eqnarray}
  \label{eq:res1}
 && T_{eff} \propto\\\nonumber
&&\propto\frac{k_b T_{eff}^{scale}}{M}-\\\nonumber
&&-<\sin^4(\theta)>_{tr}\frac{2k_bT_{eff}^{scale}m}{M^2}\frac{S_1\rho_1}{\sum S_i\rho_i\left (\frac{T_i}{T_{eff}^{scale}} \right )^{\frac{1}{2}}<sin^2(\theta)>_i}\left (\frac{T_1}{T_{eff}^{scale}}-1 \right )\left (\frac{T_{eff}^{scale}}{T_1}-\frac{1}{4} \right )+\\\nonumber
&&+<\sin^3(\theta)>^2_{tr}\frac{7\pi k_bT_{eff}^{scale}m}{8M^2}\frac{S_1^2\rho_1^2}{\left (\sum S_i\rho_i\left (\frac{T_i}{T_{eff}^{scale}} \right )^{\frac{1}{2}}<sin^2(\theta)>_i \right )^2}\left (\frac{T_1}{T_{eff}^{scale}}-1 \right )\left (\frac{T_1}{T_{eff}^{scale}}-\frac{4}{7} \right ),
\end{eqnarray}
\end{widetext}
where $<sin^n(\theta)>_{tr}$ are the form factors (\ref{eq:sinn}) of the triangle part of the motor. These form factors depends on the rotation state $\alpha$ and define the temperature difference between symmetric and asymmetric states.

Transition from symmetric to nonsymmetric motor state occurs universally near $T_1=T_2$. In this case $T_1=T_{eff}^{scale}$ and effective temperature of the motor is independent of triangle's orientation, see for example,(\ref{eq:res1}). Therefore, general form of effective temperature (\ref{eq:res1}) near point $T_1=T_2$ is:
\begin{eqnarray}
  \label{eq:Tbm1}
  T_{eff}=T_0+A_1\Delta T+A_2\Delta T\Delta\alpha,
\end{eqnarray}
where $\Delta T = T_1-T_2$ and $A_i$ are system dependent coefficients. Rotation $\Delta\alpha$ reduces effective temperature either for positive or negative $\Delta T$ depending on the sign of the coefficient $A_2$. Therefore, it is always possible to facilitate the motion by tuning temperature difference $\Delta T$.

As there are two types of transitions between symmetric and asymmetric states of Brownian motor, transition of the first type depends on the change of the form of Brownian ratchet. In this case effective temperature of the motor might reduce because of change in the form factors $<sin^4(\theta)>$, $<sin^2(\theta)>$ or perimeter $S$, see the  $<sin^4(\theta)>$ term of (\ref{eq:res1}). The motion comes as a side effect of optimization of friction coefficients (\ref{eq:gammai1}) along the axis of the motion.

Transition of the second type is driven solely by asymmetry $<sin^3(\theta)>$ and associated velocity (\ref{eq:vtriang}). For example, $<sin^4(\theta)>$ and $<sin^2(\theta)>$ of an equilateral triangle are independent of orientation $\alpha$ (see Fig. 2). Therefore, the change in effective temperature (\ref{eq:res1}) is:
\begin{widetext}
\begin{eqnarray}
  \label{eq:dteff1}
 && \Delta T_{eff} \propto\\\nonumber
&&\propto\Delta<\sin^3(\theta)>^2_{tr}\frac{7\pi k_bT_{eff}^{scale}m}{8M^2}\frac{S_1^2\rho_1^2}{\left (\sum S_i\rho_i\left (\frac{T_i}{T_{eff}^{scale}} \right )^{\frac{1}{2}}<sin^2(\theta)>_i \right )^2}\left (\frac{T_1}{T_{eff}^{scale}}-1 \right )\left (\frac{T_1}{T_{eff}^{scale}}-\frac{4}{7} \right ),  
\end{eqnarray}
\end{widetext}
Favorable conditions for transition to the asymmetric state with maximum propulsion and minimum effective temperature are:
\begin{eqnarray}
  \label{eq:trcond1}
 \frac{4}{7}<\frac{T_1}{T_{eff}^{scale}}<1,  
\end{eqnarray}
where scale temperature $T_{eff}^{scale}$ is defined by (\ref{eq:Teffsc1}). In space of temperatures $(T_1,T_2)$, these conditions are presented in Fig. 3. 

Transition of the second type results in dynamics of a general Brownian motor near $T_1\approx T_2$ as an Onsager linear equation\cite{Onsager1931}\cite{VandenBroeck2006} with coefficients that depend on stochastic process:
\begin{eqnarray}
  \label{eq:veq1}
  &&v=L_{vT}(\alpha)\frac{\Delta T}{T^2},\\\nonumber
  &&J_{\alpha}(\alpha,\Delta T_{eff}(\alpha))=0,
\end{eqnarray}
where $L_{vT}$ is Onsager coefficient, $\Delta T_{eff}(\alpha)$ reduces with development of motion (\ref{eq:dteff1}) and $J_{\alpha}$ is the flux of the system in $\alpha$ space towards a state with maximum motion ability, for example, see (\ref{eq:jtheta1}). Orientation $\alpha$ becomes general parameter that describes asymmetry and the corresponding velocity $v(\alpha)$, i.g. (\ref{eq:vtriang}). In the limit of strong coupling to the effective temperature,  $v=0$ or $v=v_{max}$ are the only possible states of the motor.
\begin{figure}
\resizebox{0.5\textwidth}{!}{\includegraphics{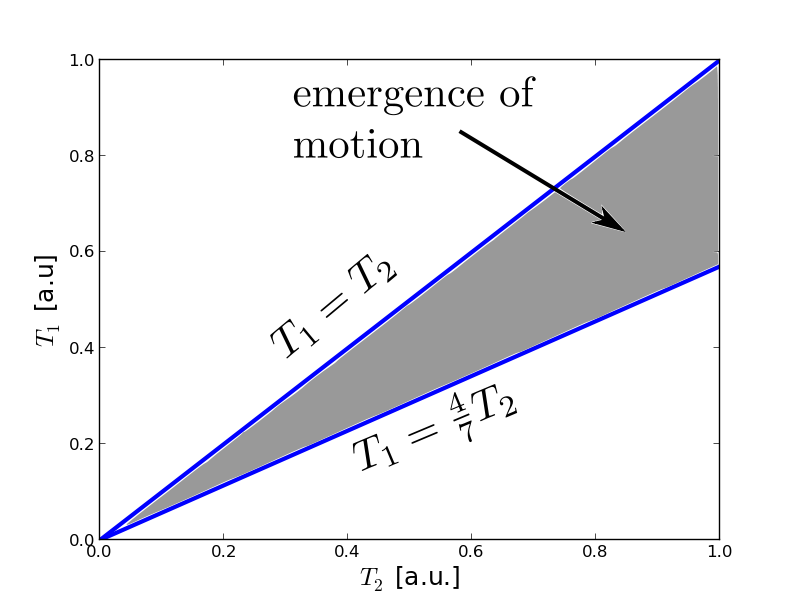}}
\caption{Phase diagram of Triangulita with rotating equilateral triangle. Self-assembly of Triangulita with equilateral triangle is an example of a system driven only by the development of motion. Self-assembly because of reduction of effective temperature in the region  $T_1\approx T_2$ is universal property of Brownian motors. Therefore, self-assembly driven by effective temperature is a general mechanism.}
\label{fig3}
\end{figure}

Generalization of (\ref{eq:veq1}) to many degrees of freedom is:
\begin{eqnarray}
  \label{eq:onsag1}
  &&J_i=L_{ij}X_j,\\\nonumber
  &&J_{Lij}(L_{ij},T_{eff}^{ij})=0,
\end{eqnarray}
The motor-likes steady state, see eqs. (\ref{eq:vtriang}) and (\ref{eq:veq1}), corresponds to constant external forces $X_j$. The system (\ref{eq:onsag1}) might possess several solutions resembling the variety of stable states in biological systems. The steady states of (\ref{eq:onsag1}) do not nesssarily hold Prigogine's principle of minimum entropy production $\partial S/\partial t = \sum J_iX_i$.

The notion of effective temperature\cite{Cugliandolo2011a} and associated transitions might be integrated in any description of a system with conformation transitions out of thermal equilibrium, for instance\cite{Rahav2008}. The parameters of the system cease to be constant and converge to the states of minimum effective temperature. This convergence might be associated with emergence of a new property such as motion or more complex activity. Effective temperature is a useful tool that already served to describe behavior and stability of an active medium\cite{Wang2011,Loi2008}.

Macro-molecule with preferential ATP hydrolysis site is a complementary system to two temperature Brownian motor, such as Triangulita or Smoluchowski-Feynman ratchet. ATP releases about $9k_bT$ during reaction with the hydrolysis site. The molecule, therefore, is coupled to high temperature $T_{high}$ thermal bath in addition to its environment temperature $T$. According to this work, these two temperatures define the effective temperature of the molecule and, under particular circumstances, drives its conformational changes toward development of motion. This sequence, with a two-temperature framework, holds for the Brownian motor models based on external noise correlated in time. Weak coupling to ATP hydrolysis might be described either by two temperatures or by external modulation\cite{Cilla2001,Magnasco1998,Gomez-Marin2006,Martinez2013,Sekimoto1997}.

The proposed mechanism for the emergence of motor abilities matches the phenomenon of chemotaxis\cite{Berg1972}. The mutation rate of a Brownian motor reduces with the development of motor abilities just as the effective diffusion coefficient of bacteria reduces toward the areas with high concentration of food. This analogy indicates generality of the mutation modulation mechanism in the early living systems.

Thus, a general mechanism favoring self-assembly of Brownian motors is predicted. This mechanism is a common property of several temperature systems near thermal equilibrium. The steady state of a Brownian motor is described by the nonlinear stochastic version of Onsager relations. It could be a novel description of molecular motors and the future artificial nanomachines with complex robotic functions. These results open a new way to consider emergence and stability of living-like systems and to accomplish the similar phenomena using alternative technologies.

%

\end{document}